# The puzzling high velocity G5 supergiant star HD 179821 : new insight from Gaia DR2 data

M. Parthasarathy[1] , G. Jasniewicz[2] , and F. Thévenin[3]

1 Indian Institute of Astrophysics, Bangalore 560034, India e-mail: m-partha@hotmail.com

2 UMR 5299 LUPM CNRS/Université Montpellier, Place EugèneBataillon, CC 72, 34095 Montpellier Cedex 05, France e-mail: gerard.jasniewicz@umontpellier.fr

3 Observatoire de la Côte d'Azur, UMR 6202 Cassiopée / CNRS, B.P. 4229, 06304 Nice Cedex 4, France e-mail: Frederic.thevenin@oca.eu

**ABSTRACT**

HD 179821 is classified as G5 Ia star. From the IRAS colors and spectral energy distributions it was classified as a post-AGB star. But some studies classify it as a massive (30 to 19 $M_{sun}$) post-red supergiant evolving to become a Type II supernova. Its mass and evolutionary status remained a hotly debated question even after several detailed spectroscopic studies as the distance was not known. We use the parallax of HD 179821 from the second Gaia data release, and deduce its distance 2959 ± 409 pc and its absolute magnitude MV = -5.7 ± 0.3. Using the absolute magnitude determined in this paper, we show that HD 179821 fits very well with post-AGB tracks in the H-R diagram. Our results clearly confirm that HD 179821 is a post-AGB star of mass in the range of 0.8 $M_{sun}$. It is not a 30 $M_{sun}$ red supergiant. The progenitor mass of this post-AGB star may be in the range of 4 $M_{sun}$ but may be a bit more.

## 1. Introduction

HD 179821 (IRAS 19114+0002) was classified as a post-AGB star by Pottasch & Parthasarathy (1988) based on IRAS colors and SED. Its IRAS colours and SED are similar to that of high galactic latitude post-AGB star HD 161796 (Parthasarathy & Pottasch 1986). It was also classified as a massive (30 $M_{sun}$) red supergiant, post-red supergiant evolving to become a Type II supernova (Zuckerman & Dyck 1986, Jura & Werner 1999, Jura et al. 2001). Zuckerman & Dyck (1986) from CO observations estimated the distance of HD 179821 to be 6kpc. Adopting this 6kpc distance Hawkins et al. (1995) and Kastner & Weintraub (1995) also considered to be a very massive star with massive circumstellar shell.

The spectral type of HD 179821 is G5 Ia (Keenan 1983; Keenan & McNeil 1989; Buscombe 1984; Hrivnak et al. 1989) and galactic latitude b = -4.96°. Zacs et al. (1996) derived the Radial Velocity (RV) of the star from high resolution spectra to be +88.4 km s$^{-1}$ which shows that it is a high velocity star.

On the other hand, the work of Josselin & Lèbre (2001) presents high sensitivity radio observations of HD 179821 in the circumstellar $^{12}$CO and $^{13}$CO. They detected HCO+ and found $^{12}$CO/$^{13}$CO < 5 ± 1. Such a low value is characteristic of (post-)AGB stars with low or intermediate-mass progenitors, and not of a 30 M$_{sun}$ red supergiant.

Ferguson & Ueta (2010) made differential proper motion study of the circumstellar shell based on HST observations. They concluded that it is a post-AGB star and estimated distance of 3.91 ± 3.23 kpc.

Recently Ikonnikova et al. (2018) presented the results of multicolor photometry obtained during 2009 to 2017. They found that HD 179821 displays low amplitude 0.2 magnitude semi-periodic brightness variations, and concluded that the star's behaviour is similar to that of yellow hypergiants located near the very unstable Yellow Void, and differs significantly from ordinary post-AGB objects.

Thus HD 179821 is either classified as a post-AGB star descended from a low-mass progenitor or as a very massive red supergiant. This research note follows a previous paper (Thévenin et al. 2000) in which we concluded that HD 179821 is a low mass post-AGB star with $T_{eff}$ = 5660 K and not a massive red supergiant. In Section 2 of this paper we report accurate distance and $M_V$ based on the analysis of the accurate parallax of HD 179821 from the second Gaia data release (hereafter GDR2; Gaia Collaboration 2018). In Section 3 we discuss the new data, relate recent studies and give our conclusions in Sect.4.

## 2. New observational data from the second Gaia data release

The GDR2 parallax of HD 179821 is found to be 0.310239 ± 0.051225 mas (Gaia Collaboration 2018), giving a parallax precision about 16%. The distance of the star is thus approximately 3000 pc, but according to Bailer-Jones et al. (2018) going from a Gaia parallax to a distance is a non-trivial issue and cannot be obtained by simply inverting the parallax. In the following we adopt the distance given by the inference procedure developed by Bailer-Jones et al. (2018) : 2959 ± 409pc. Using this distance and observed V magnitude (8.19) and B−V color (1.509) from Høg et al. (2000) and G5Ia spectral type and intrinsic (B−V=1.02) color of a G5Ia star and $A_V$ = 3.1 E(B−V) we find $A_V$ = 1.50 ± 0.06 and $M_V$ = −5.7 ± 0.3.

The observed (B−V) color of HD 179821 consists of interstellar and circumstellar reddening which is taken into account when we calculate $A_V$ using the E(B−V). Le Coroller et al. (2003) and Arkhipova et al. (2009) monitored the photometric variability for many years and found small amplitude light variations similar to that found in other post-AGB supergiants. They derived E(B−V) = 0.7, which is compatible with the E(B−V) value we have adopted here.

From the Gaia parallax we further deduce that HD 179821 is approximately z = -260 pc below the galactic plane. The large $V_{LSR}$ = 105 kms$^{-1}$ observed by Zuckerman & Dyck (1986) give the central velocity of the CO or HCN profile with respect to the local standard of rest. The 6 kpc distance being no more reliable, the large VLSR is not due to differential galactic rotation. HD 179821 is simply a high velocity star, confirmed by the Gaia RVS value +81.8 ± 3.7kms$^{-1}$ (Sartoretti et al. 2018).

## 3. Discussion

The derived $M_V$ = -5.7 clearly establishes that HD 179821 is not a 30 or 19 $M_{sun}$ post- red supergiant. Indeed, for such a mass Ekström et al. (2012) give a luminosity $\log(L/L_{sun})$ = 5.5, which corresponds to $M_{bol}$ = -9.0. Hereafter we take BC = -0.33 for the bolometric correction for a G5 supergiant, and thus $M_{bol}$ = -6.0 ± 0.4 and $\log(L/L_{sun})$ = 4.3 ± 0.2 for HD 179821.

Concerning the effective temperature of HD 179821 there is no consensus among researchers. However we would like to emphasize here that there is not large error or uncertainty in the MK spectral type assigned to this star by Keenan (1983), Keenan & McNeil (1989), Buscombe (1984), and Hrivnak et al. (1989). These authors have classified the spectrum of HD 179821 as G5 I, and never as that of a F0Ia-F2Ia (7350K) star. Sahin et al. (2016) derive $T_{eff}$ = 7350K which is not consistent with its observed spectral type and B-V color. The spectroscopic studies by Zacs et al. (1996), Reddy & Hrivnak (1999), Thévenin et al. (2000), Kipper (2008) and Sahin et al. (2016) yielded very different $T_{eff}$ values ranging from 5500 K to 7350 K. In the following we adopt the effective temperature 5660 K (log $T_{eff}$=3.75) according to Thévenin et al. (2000), in agreement with the G5 MK spectral classification, to show its position in the H-R diagram. We have not used the $T_{eff}$ values in our calculations. We used only the observed V magnitude, observed B-V colour and observed spectral type. In the HR diagram the post-AGB stars evolutionary track is towards the left from the tip of the AGB and with constant luminosity even though their $T_{eff}$ varies as they evolve towards hot post-AGB stars and early stages of young planetary nebulae.

In order to study the evolutionary status of HD 179821 in the H-R diagram, we use the post-AGB evolutionary models from Miller Bertolami (2016) which are based on outdated micro- and macro-physics. The models are computed for initial masses between 0.8 and 4 $M_{sun}$ and for a wide range of initial metallicities, the solar metallicity being Z = 0.02. They include all previous evolutionary stages from the Zero Age Main Sequence to the White Dwarf phase. The evolutionary sequences in Fig. 1 correspond to the sequence presented in Table 3 of the article of Miller Bertolami. These new models are 0.1 to 0.3 dex brighter than the old post-AGB stellar evolution models with similar remnant masses. The final masses are in the relevant range for the formation of planetary nebulae with central stars masses in the range 0.5 to 0.8 $M_{sun}$.

The location of HD 179821 indicates that the progenitor of this post-AGB star may have an initial main sequence mass of 4 ± 1 $M_{sun}$. We can call it as a super post-AGB star resulting from an intermediate mass AGB star or from a star which is in the lower limit of a super-AGB star.

Sahin et al. (2016) used the very broad and very strong Na D absorption feature in the high resolution spectrum of HD 179821. This Na D feature is strongly blended with stellar, interstellar and circumstellar Na D lines and it is impossible to disentangle the interstellar and circumstellar Na D lines to estimate reliable E(B-V) value. Considering the uncertainties and errors in the E(B-V) value they estimated it is similar to the value that we adopted here. In deriving the 19 $M_{sun}$ mass of HD 179821 Sahin et al. (2016) used the unreliable $M_V$ value derived from the equivalent width of OI 7774 Å triplet feature and spectroscopic surface gravity log g. The post-AGB supergiants and normal massive supergiants have similar spectroscopic surface gravity values. The spectroscopic surface gravity is not a reliable parameter to derive mass of a supergiant. Therefore the 19 $M_{sun}$ for HD 179821 derived by them is unreliable and not correct.

## 4. Conclusions

Based on the accuracy of the parallax of HD 179821 taken from the GDR2, and our confidence in the evolutionary tracks of post-AGB (Miller Bertolami 2016) and supergiants (Ekström et al. 2012) we definitively conclude after 30 years of debate, that HD 179821 is a post-AGB star of mass in the range of 0.8 $M_{sun}$. The bolometric magnitude clearly excludes the possibility, defended by many authors, that it is a very massive (30 to 19 $M_{sun}$) post-red supergiant star.

*Acknowledgements.* This study has made use of the SIMBAD database, the VizieR catalogue access tool and of NASA's Astrophysics Data System, operated at CDS, Strasbourg, France. The authors are grateful to Olivia Mancuso for her review of English phrasing and grammar in this article.

Fig. 1. HR diagram. HD 179821 is located with $T_{eff}$ = 5660K and $\log(L/L_{sun})$ = 4.3 ± 0.2 (see Sect.3). The post-AGB models come from Miller Bertolami (2016) ; the initial masses and metallicities are indicated.

Figure with evolutionary tracks on log(L/L_sun) vs log(Teff) diagram showing HD 179821 position. Legend: $M_{init} = 3 M_{sun}$; $Z = 0.01$ (black), $M_{init} = 3 M_{sun}$; $Z = 0.02$ (blue), $M_{init} = 4 M_{sun}$; $Z = 0.02$ (red).